# Centrifugal Experiments with Simulated Regolith: Effects of Gravity, Size Distribution, and Particle Shape on Porosity

By Tomomi OMURA[1)], Masato KIUCHI[1)], Carsten GÜTTLER[2)], Akiko M. NAKAMURA[1)]

[1)] *Kobe University, Kobe, Japan*

[2)] *Max-Planck-Institute for Solar System Research, Göttingen, Germany*

Porosity is a key characteristic of the regolith on the surface of small bodies. The porosity of the regolith on the surface of asteroids is changed by applied pressure, and the relationship between pressure and porosity depends on the particle properties of the regolith. We performed compression measurements on samples of different materials, particle size distributions, and shapes to examine the relationship between particle size and pressure required for compression. We used a centrifuge and a compression testing machine for the experiments. The applied pressure for the centrifuge and the compression testing machine experiments ranged from $10^2$ to $5\times10^3$ Pa and from $10^4$ to $5\times10^6$ Pa, respectively. The initial porosity before compression was generally higher for samples with smaller particles and narrower particle size distributions. A sample compressed more easily when it consisted of smaller particles, probably due to smaller frictional forces between particles. We estimated the porosity of granular asteroids based on the results of our experiments. The estimated porosity at the center of a homogeneous asteroid with a radius of 50 km is 0.43. This porosity is typical of the porosity of asteroids of similar size.

## Nomenclature

| | | |
|---|---|---|
| $p$ | : | porosity |
| $\rho_p$ | : | particle density |
| $r$ | : | particle radius |
| $g$ | : | gravity |
| $A$ | : | Hamaker constant |
| $s$ | : | cleanliness ratio |
| $p_0$ | : | constant |
| $m$ | : | constant |
| $n$ | : | constant |
| $\Omega$ | : | diameter of $O^{-2}$ ion |
| $\mu$ | : | reduced shear modulus |
| $a_0$ | : | radius of contact area |
| $G$ | : | gravitational constant |
| $\rho$ | : | bulk density of asteroid |

## 1. Introduction

Both recent and expected near-future spacecraft exploration of asteroids has included landing on and sampling from the surface. The surfaces of asteroids are known to be covered by regolith, and the porosity of this regolith is an important characteristic affecting its physical properties, such as strength of the surface against meteoroid impacts or spacecraft landings.[1)] The porosity of a granular bed consisting of particles of fine dust has been shown to decrease with the pressure applied to the sample in laboratory experiments.[2,3)] A similar phenomenon would occur in regolith on the surface of an asteroid.

Models of the relationship between pressure and porosity of very fluffy aggregates of dust in protoplanetary disks have been developed.[4)] However, these models may not be directly applicable to regolith layers because the porosity of the aggregate in the models is larger than 0.9, too large to be applied to regolith. The porosities of regolith layers are generally lower than those of dust aggregates in protoplanetary disks. Therefore, it is expected that the physical process of compaction is different.

Porosity is related to the coordination number of the constituent particles in particle layers, which increases as porosity decreases; thus, the force required to deform particle layers changes with coordination number.[5)] When particle layers have high porosity and low coordination numbers, particles are rearranged by rolling because the energy required to roll particles is smaller than the energy required to slide them. However, when porosity is low and the coordination number exceeds 6, particles are rearranged by sliding because they are locked in triaxial directions. The porosity of a particle layer at this transition has been shown to be ~0.7.[5)] The bulk porosity of regolith on small bodies has been estimated to be about 0.4-0.9.[6,7)] Therefore, the sliding force between particles can be expected to be relevant in the rearrangement of regolith particles and the compression of the regolith layer.

In this study, we investigate the relationship between pressure and the porosity of simulated regolith using different materials and different material size distributions, and then estimate the porosity of granular asteroids based on our experimental results.

## 2. Experiments

### 2.1. Samples

The powder samples used in this study are listed in Table 1. We used silica sand 1-3 of irregular shape and with different size distributions, fly ash of spherical shape, fused alumina 1-3 of irregular shape and with different size distributions, and basalt fragments smaller than 210 μm prepared by an impact experiment (hereafter called "basalt"). The size distributions of the samples, shown in Fig. 1, were determined by laser diffractometry. We sieved the samples into a cylindrical container with a diameter of 5.8 cm and depth of 3.3 cm (for basalt, diameter of 2.7 cm and depth of 1.4 cm) and then leveled off the part of the bed exceeding the height of the container with a spatula. All experimental manipulations were conducted under atmospheric pressure. The porosity of the samples was determined by the mass weighed and volume measured after compression. The volumes of the samples were determined from their heights.

Table 1. Sample description.

| Sample | Density (g/cm$^3$) | Median diameter (μm) | Shape |
|---|---|---|---|
| Silica sand 1 | 2.5 | 13 | irregular |
| Silica sand 2 | 2.5 | 19 | irregular |
| Silica sand 3 | 2.5 | 73 | irregular |
| Fly ash | 2.0 | 4.8 | spherical |
| Fused alumina 1 | 4.0 | 4.5 | irregular |
| Fused alumina 2 | 4.0 | 23 | irregular |
| Fused alumina 3 | 4.0 | 77 | irregular |
| Basalt | 2.7 | 53 | irregular |

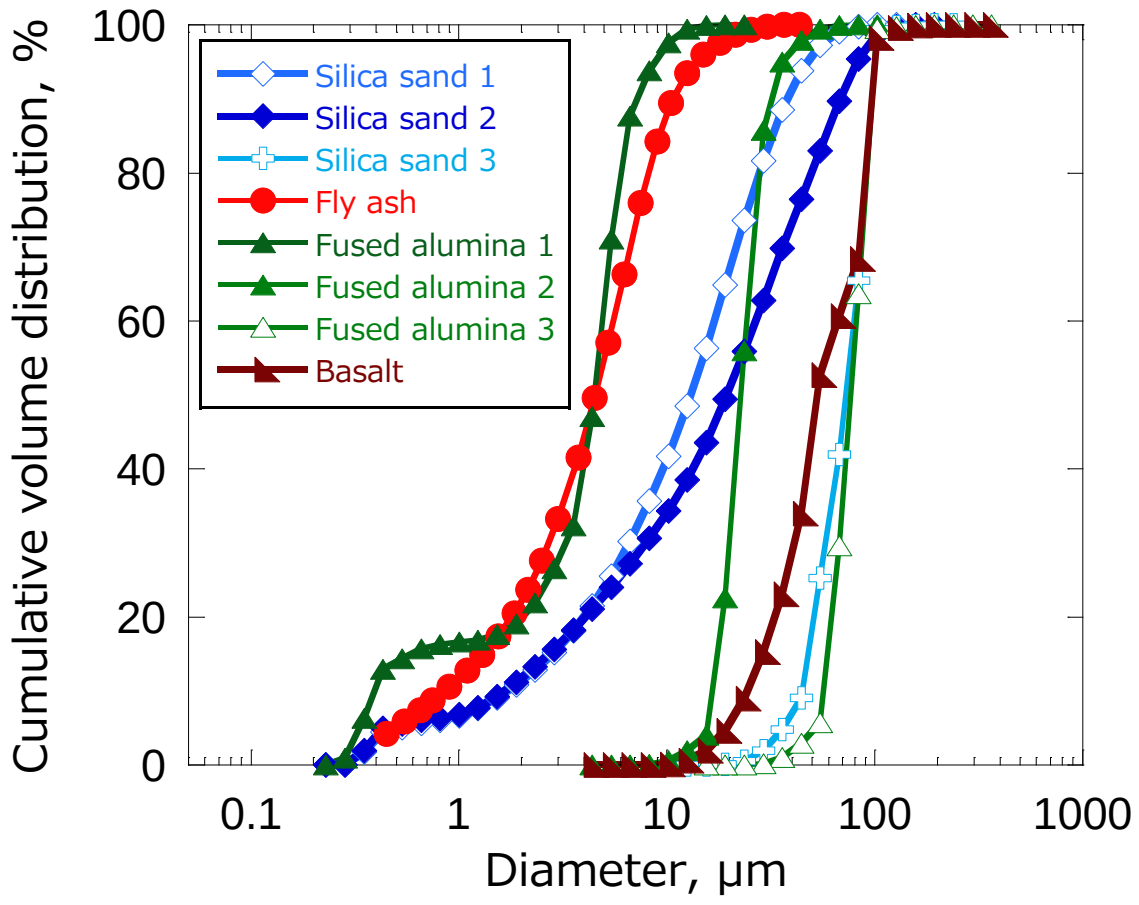

Fig. 1. Samples particle size distribution.

### 2.2. Experimental methods

Compression of the samples was performed using a centrifuge and a compression testing machine at Kobe University. Figures 1 and 2 show schematic views of the experimental setups. Compression by centrifuge simulates self-gravitational compression. However, we could not achieve high pressure with the centrifuge used in this study. Therefore, we used a compression testing machine for the high-pressure range.

#### 2.2.1. Centrifuge

The centrifuge consisted of two ends. An experimental arrangement was mounted on one end and a personal computer (PC) for the recording of data was mounted on the opposite end. A counter balance was mounted on the PC side. Images of the surface of each sample were recorded during each experiment by a camera installed on the experimental apparatus. The images were saved to the hard disc of the PC and checked after the power supply of the centrifuge had been turned off. The simulated gravity for the samples resulted from the centrifugal force and the gravity of Earth. We could change the simulated gravity within 1-17 times Earth gravity. The rotation velocity was changed manually changing the voltage applied to the motor using a dial. The rotation velocity was determined precisely by a sensor consisting of a bar code mounted near the axis of rotation and a bar code sensor. The sensor detected the intensity of light reflected from the bar code, and returned a corresponding voltage. We could measure the time necessary for the sensor to pass over the bar code. In this study, the centrifugal acceleration at the middle depth of the sample container was calculated and then converted to overburden pressure.

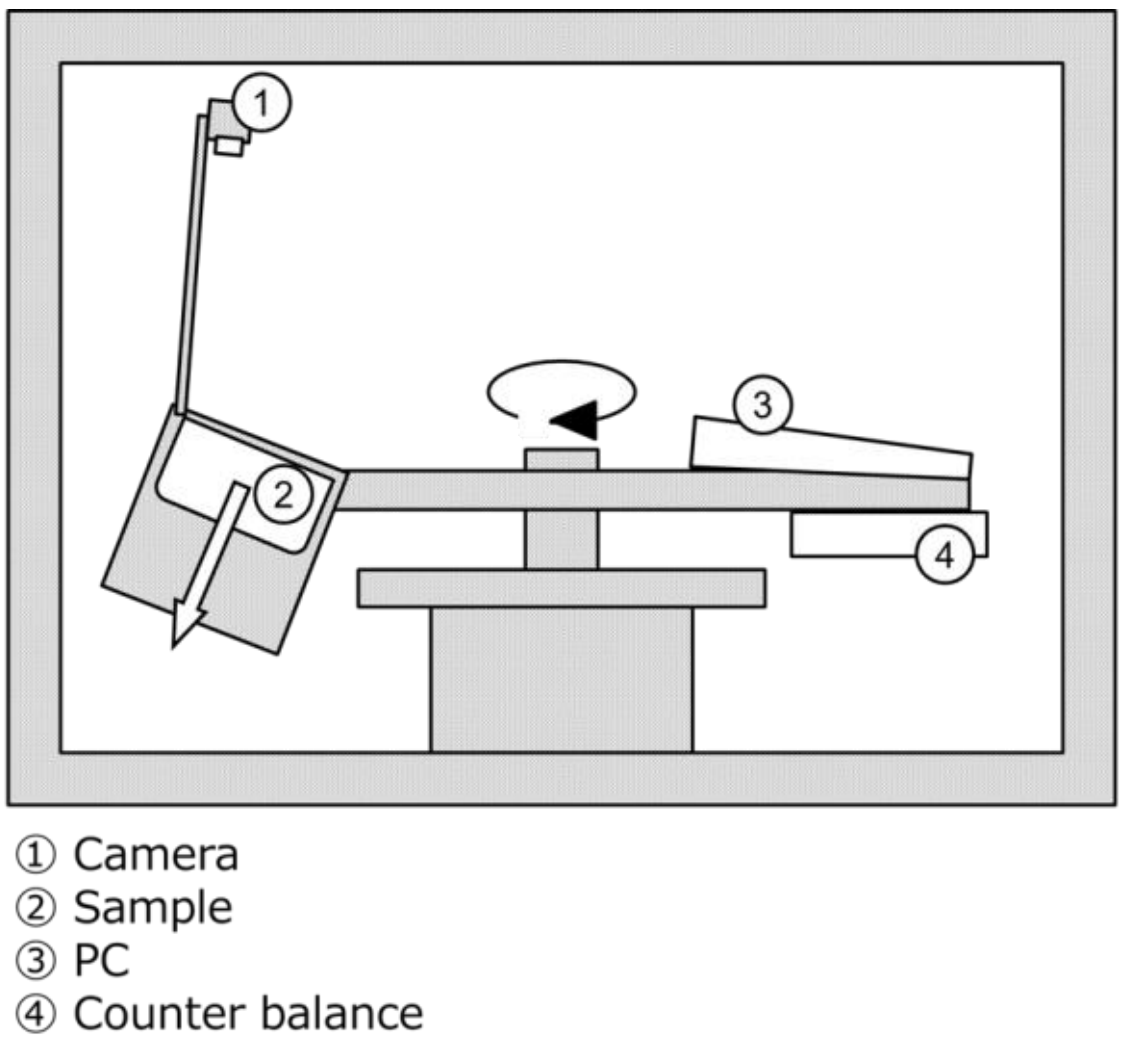

Fig. 2. Centrifuge.

#### 2.2.2. Compression testing machine

The sample container and a compression piston were placed between the pressure plates of a compression testing machine. We used cylindrical containers with inner diameter of 58 mm and height of 33 mm. The diameter of the piston was slightly smaller than the inner diameter of the container. Before beginning the compression, the bottom of the piston was lowered into the top of the container. That is, the piston was lowered into the container to ensure alignment between the piston and the wall of the container. For silica sand 3 and fused alumina 2 and 3, the starting point of the piston was 1 mm below the top of the container; for the other samples, it was 2 mm lower. The loading velocity was 0.6 mm/min. During the measurements, downward displacement and compressive force were recorded by the PC at 1 s intervals. The pressure acting on the sample was calculated by dividing the compressive force by the cross-sectional area of the piston. The volume of the sample was estimated from the height of the sample, which was calculated by the displacement of the piston. We did not conduct this measurement for basalt powders because of the small volume of the sample.

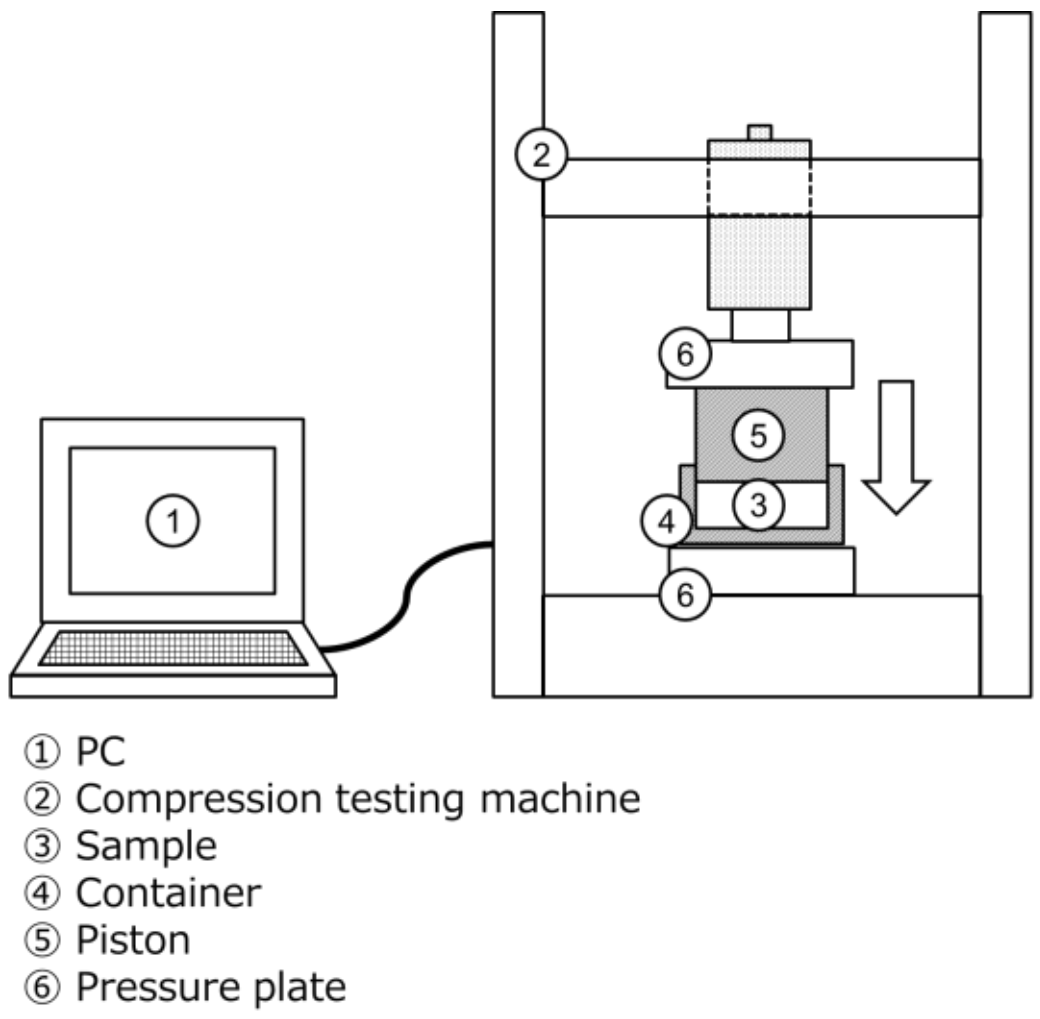

Fig. 3. Setups of the compression testing machine.

## 3. Results and Discussion

### 3.1. Porosity before compression

Figure 4 shows the relationship between the median diameter of sample constituents and the porosity before compression. The error bars represent the standard deviation of the porosity measurements. Samples with the same composition tended to have higher porosity when the grain size was smaller. The relationship between particle size and porosity is expressed by the following formula:[7)]

$$p = p_0 + (1 - p_0)\exp\left\{ -m\left( \frac{As^2}{64\pi\Omega^2 \rho_p g r^2} \right)^{-n} \right\}. \quad (1)$$

The results of Eq. (1) are also shown in Fig. 4. The blue solid, red solid, green dashed, and brown dashed curves are derived from Eq. (1) based on the density for each sample and the Hamaker constants for $SiO_2$ (silica sand and fly ash), $Al_2O_3$, and basalt, respectively. The porosities of the silica sand particles were roughly consistent with those expected from Eq. (1), especially the porosities of silica sands 1 and 3. The porosity of silica sand 2 was slightly lower than that expected from Eq. (1), presumably because this sand had the widest size distribution among the three silica sands. The porosities of the fused alumina were in agreement with or slightly higher than values estimated from Eq. (1). The porosities of the fly ash and the basalt fragments were slightly lower than expected based on Eq. (1), presumably because void space was filled by small particles.

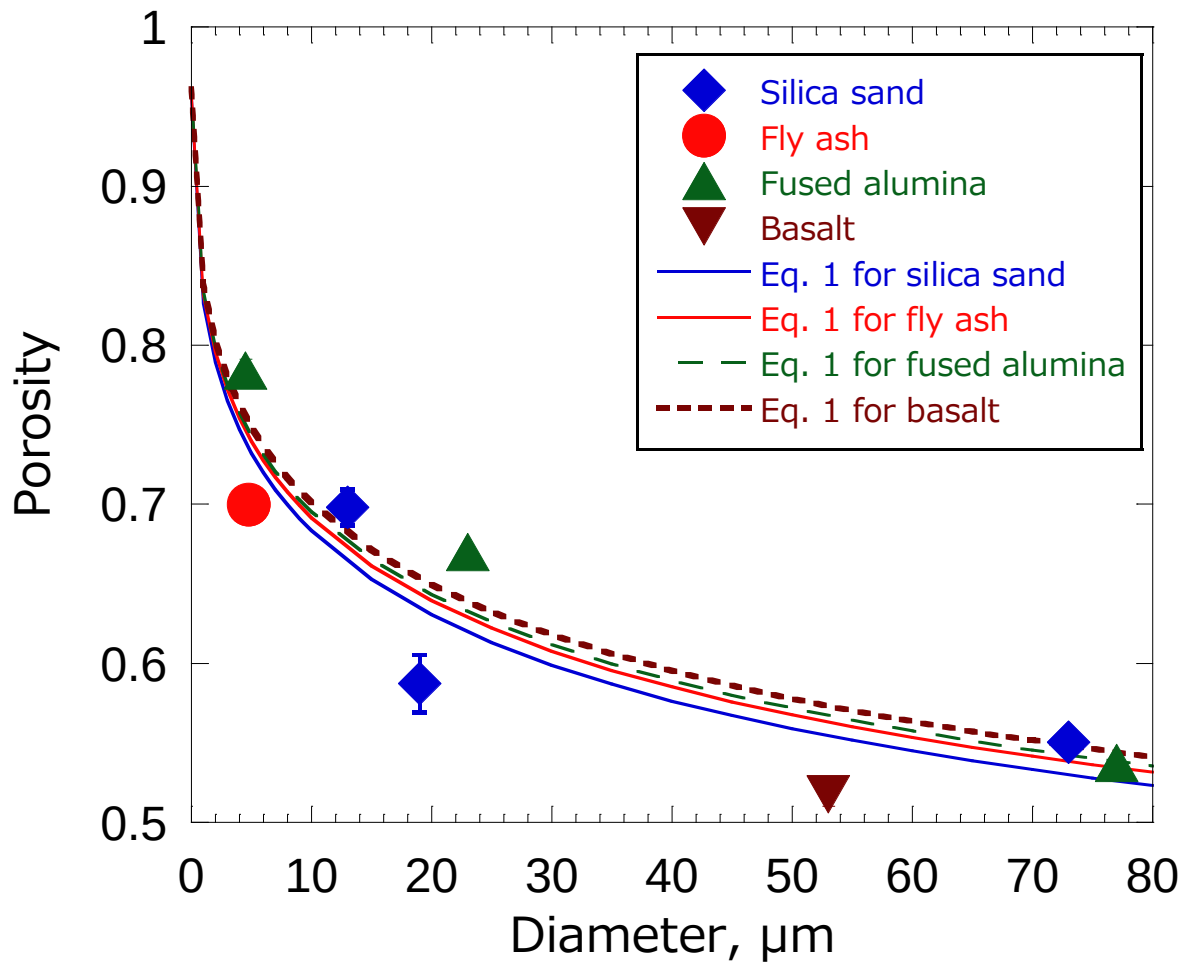

Fig. 4. Porosity before compression.

### 3.2 Porosity after compression

Figure 5 shows the results of the centrifuge and compression testing machine experiments. The results of the centrifuge experiments are shown by symbols, and the results of the compression testing machine experiments are shown by curves. The errors in Fig. 5, defined by errors in sample volume, are smaller than the symbols or the thickness of the curves. Although the pressure applied by the centrifuge was volume pressure, which is somewhat different from piston compression, the results of the centrifuge and compressive testing machine experiments were roughly consistent, i.e., they showed on similar curves. This suggests that the overburden pressure, or gravitational pressure, applied by the centrifuge had a similar effect on the samples to that applied by the piston of the compression testing machine.

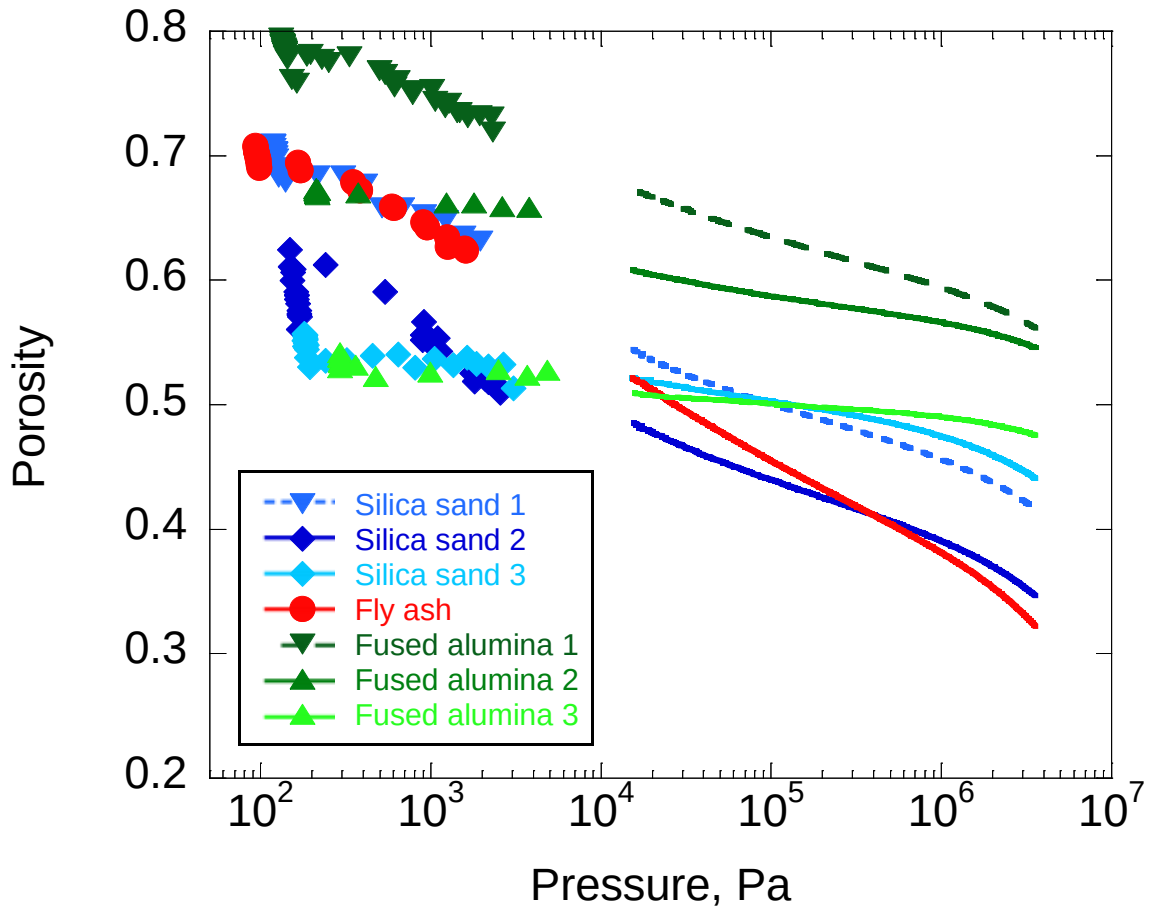

Fig.5. Results of compression experiments.

The porosities of different samples differed even when the pressure applied to the samples was the same. We fit logarithmic functions to the data of Fig. 5. Note that the slope of the curve in Fig. 5 becomes steeper for those samples that were more easily compressed. Figure 6 shows the relationship between the slope of the curves in Fig. 5 and the frictional force between the particles. Errors defined by fitting deviation are smaller than the symbols in Fig. 6. Frictional force, the force required to slide particles over one another, is given by the following formula: [8)]

$$F_{fric} = \frac{\mu a_0^{\ 2}}{2\pi}. \qquad (2)$$

We assumed spherical particles and calculated the contact radius using the median diameter of the sample particles. The figure shows a systematic trend between the frictional force and the slope of the curves. Sample compression become more difficult as frictional force increased. Fly ash seemed to be compressed more easily than the other samples due to the smaller frictional force calculated from Eq. (2), and possibly also due to its spherical shape.

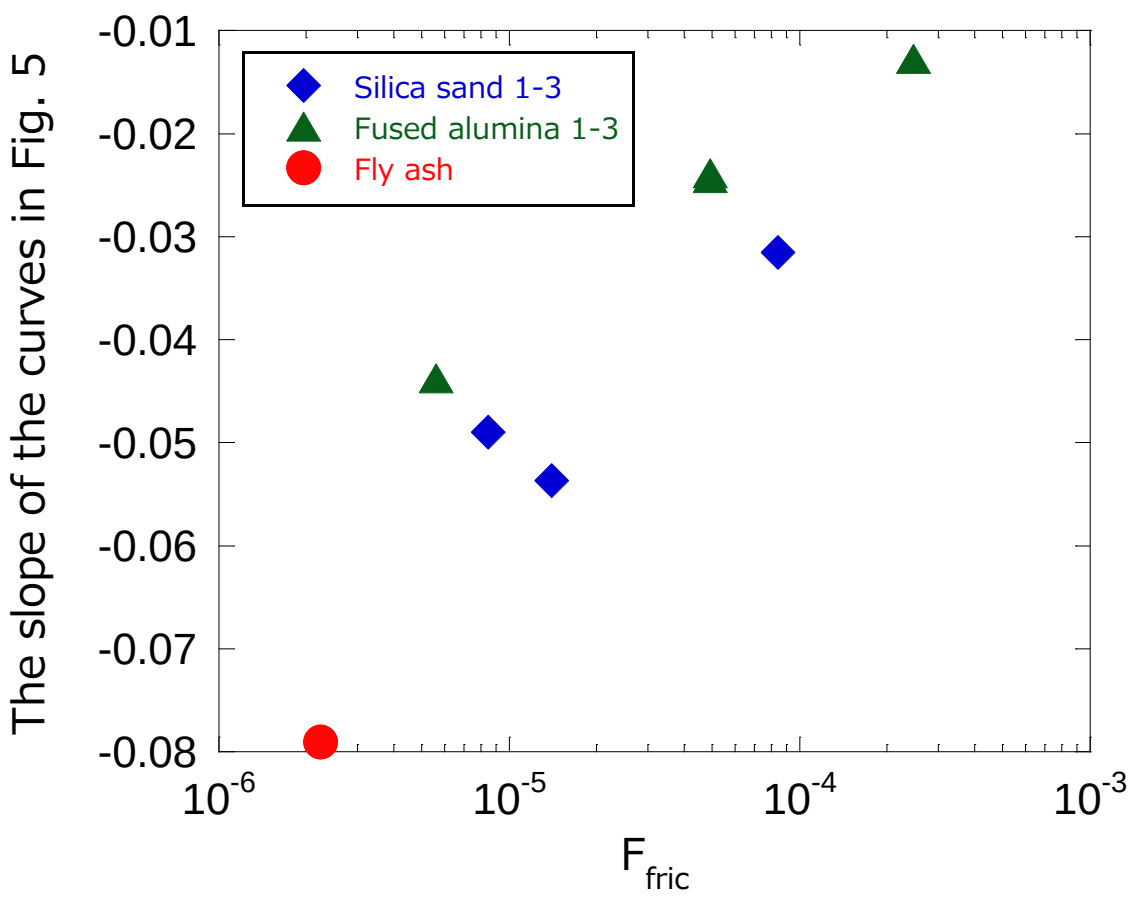

Fig. 6. Relationship between the slope of the curve in Fig. 5 and the frictional force.

### 3.3. Compression by self-gravity

The interiors of granular asteroids are subject to pressure from the soil due to self-gravity. We calculated the pressure at the center of an asteroid of radius *R* and determined the porosity using our experimental results. We assumed that the asteroid consists of regolith-sized particles. The pressure at the center of a homogeneous asteroid is given by the following formula:

$$P(0) = \frac{2}{3}\pi\rho^2 G R^2. \qquad (3)$$

The porosity of a granular asteroid with a radius of 50 km and consisting of particles similar to the silica sand 1 particles would be 0.43. This porosity is typical of the observed porosity of similar-sized asteroids. However, based on thermal observations, the grain size of the regolith on the surface of Lutetia (diameter of 95.8 km) is estimated to be 210 μm,[6)] which is about 20 times larger than the median diameter of silica sand 1. We note that our experiment was performed in air and that the compressibility may be different in vacuum. We also note that this application is valid for only small bodies. Large bodies would have a density gradient due to self-gravity throughout the interior of the body.

## 4. Summary

We performed compression experiments with granular samples of different materials, grain shapes, and size distributions. The porosity of the samples before compression tended to be higher when the median grain diameter was smaller. Our results were basically consistent with an empirical formula presented in the previous study. However, the particle samples with a narrow size distribution had a slightly higher porosity than the value expected from the empirical formula. The overburden pressure applied by centrifugal acceleration (one of our methods) or gravitational pressure had similar effects to the pressure exerted by compression with a piston (our second method). Different samples had different porosities, even if the same pressure was applied. Furthermore, the force required for compression of the sample increased as the particle diameter increased, possibly due to increased friction force between particles. The surfaces of large asteroids are covered with smaller particles, which may therefore be compressed more easily.

We estimated the porosity of granular asteroids based on our experimental results, and the estimated porosity was consistent with the observed porosity of similar-sized asteroids. However, we note that our experiment was performed in air. Compression of regolith particles may be more difficult in vacuum.